\documentclass[letterpaper]{emulateapj}
\singlespace

\newcommand{\WMAP}{\textsl{WMAP}}
\newcommand{\map}    {{\sl WMAP}}
\newcommand{\cobe}   {{\sl COBE}}

\newcommand{\be}{\begin{equation}}
\newcommand{\ee}{\end{equation}}
\newcommand{\ba}{\begin{eqnarray}}
\newcommand{\ea}{\end{eqnarray}}
 
\newcommand{\mb}{\mathbf}

\newcommand{\bg}[1]{\mbox{\boldmath$#1$}}
\newcommand{\bgb}{\bg{\beta}}
\newcommand{\bga}{\bg{\alpha}}

\def\ltsima{$\; \buildrel < \over \sim \;$}
\def\ltsim{\lower.5ex\hbox{\ltsima}}
\def\gtsima{$\; \buildrel > \over \sim \;$}
\def\gtsim{\lower.5ex\hbox{\gtsima}}

\newcommand{\nside}     {\mbox{$N_{\rm side}$}} 

\shorttitle{\WMAP\ 5-year Polarized Map Estimation}
\shortauthors{Dunkley et al.}

\slugcomment{Astrophysical Journal Accepted}

\begin{document}

\title{Five-Year Wilkinson Microwave Anisotropy Probe 
(WMAP\altaffilmark{1}) Observations: Bayesian Estimation of CMB Polarization maps}

\author{
{J. Dunkley} \altaffilmark{2,3,4}, 
D. N. Spergel \altaffilmark{3,5}, 
E. Komatsu \altaffilmark{6}, 
G. Hinshaw \altaffilmark{9}, 
D. Larson \altaffilmark{8},
M. R. Nolta \altaffilmark{7}, 
N. Odegard \altaffilmark{10},
L. Page \altaffilmark{2}, 
C. L. Bennett \altaffilmark{8},  
B. Gold \altaffilmark{8}, 
R. S. Hill \altaffilmark{10},
N. Jarosik \altaffilmark{2}, 
J. L. Weiland \altaffilmark{10},
M. Halpern \altaffilmark{11}, 
A. Kogut \altaffilmark{9}, 
M. Limon \altaffilmark{12},
S. S. Meyer \altaffilmark{13},
G. S. Tucker \altaffilmark{14}, 
E. Wollack \altaffilmark{9}, 
E. L. Wright \altaffilmark{15}}

\altaffiltext{1}{\map\ is the result of a partnership between Princeton 
                 University and NASA's Goddard Space Flight Center. Scientific 
		 guidance is provided by the \map\ Science Team.}

\altaffiltext{2}{Dept. of Physics, Jadwin Hall, %
            Princeton University, Princeton, NJ 08544-0708}
\altaffiltext{3}{Dept. of Astrophysical Sciences, %
            Peyton Hall, Princeton University, Princeton, NJ 08544-1001}
\altaffiltext{4}{Astrophysics, University of Oxford, %
            Keble Road, Oxford, OX1 3RH, UK}
\altaffiltext{5}{Princeton Center for Theoretical Physics, %
            Princeton University, Princeton, NJ 08544}
\altaffiltext{6}{Univ. of Texas, Austin, Dept. of Astronomy, %
            2511 Speedway, RLM 15.306, Austin, TX 78712}
\altaffiltext{7}{Canadian Institute for Theoretical Astrophysics, %
            60 St. George St, University of Toronto, %
	    Toronto, ON  Canada M5S 3H8}
\altaffiltext{8}{Dept. of Physics \& Astronomy, %
            The Johns Hopkins University, 3400 N. Charles St., %
	    Baltimore, MD  21218-2686}
\altaffiltext{9}{Code 665, NASA/Goddard Space Flight Center, %
            Greenbelt, MD 20771}
\altaffiltext{10}{Adnet Systems, Inc., %
            7515 Mission Dr., Suite A1C1 Lanham, Maryland 20706}
\altaffiltext{11}{Dept. of Physics and Astronomy, University of %
            British Columbia, Vancouver, BC  Canada V6T 1Z1}
\altaffiltext{12}{Columbia Astrophysics Laboratory, %
            550 W. 120th St., Mail Code 5247, New York, NY  10027-6902}
\altaffiltext{13}{Depts. of Astrophysics and Physics, KICP and EFI, %
            University of Chicago, Chicago, IL 60637}
\altaffiltext{14}{Dept. of Physics, Brown University, %
            182 Hope St., Providence, RI 02912-1843}
\altaffiltext{15}{PAB 3-909, UCLA Physics \& Astronomy, PO Box 951547, %
            Los Angeles, CA 90095--1547}

\email{j.dunkley@physics.ox.ac.uk}

\begin{abstract}
We describe a sampling method to estimate the polarized CMB signal 
from observed maps of the sky. We use a Metropolis-within-Gibbs 
algorithm to estimate the polarized CMB map, containing Q and U Stokes 
parameters at each pixel, and its covariance matrix. These can
be used as inputs for cosmological analyses. 
The polarized sky signal is parameterized as the sum of three components: CMB, 
synchrotron emission, and thermal dust emission. The polarized Galactic 
components are modeled with spatially varying power law spectral 
indices for the synchrotron, and a fixed power law for the dust, and 
their component maps are estimated as by-products. We apply the method 
to simulated low resolution maps with pixels of side 7.2 
degrees, using diagonal and full noise realizations drawn from the \map\ 
noise matrices. The CMB maps are recovered with goodness 
of fit consistent with errors. Computing the likelihood of the E-mode 
power in the maps as a function of optical depth to reionization, $\tau$, 
for fixed temperature anisotropy power, 
we recover $\tau=0.091\pm0.019$ for a simulation with input $\tau=0.1$,
and mean $\tau=0.098$ averaged over 10 simulations. 
A `null' simulation with no polarized CMB signal has maximum likelihood 
consistent with $\tau=0$. 
The method is applied to the five-year \map\ data,
using the K, Ka, Q and V channels. We 
find $\tau=0.090\pm0.019$, compared to $\tau=0.086\pm0.016$ 
from the template-cleaned maps used in the primary \map\ analysis. 
The synchrotron spectral index, $\beta$, averaged over high signal-to-noise 
pixels with standard deviation $\sigma(\beta)<0.25$, but excluding 
$\sim 6\%$ of the sky masked in the Galactic plane, 
is $-3.03\pm0.04$. This estimate does not vary significantly with 
Galactic latitude, although includes an informative 
prior.
\end{abstract}

\section{Introduction}
\label{sec:intro}

The {\it Wilkinson Microwave Anisotropy Probe} (\map) has mapped the sky in 
five frequency bands between 23 and 94 GHz. Measurements of the 
temperature anisotropy in the CMB have led to the establishment of the 
$\Lambda$CDM cosmological model. Anisotropies in the CMB polarization at 
large scales inform us about the ionization 
history of the universe, allow us to probe a possible 
signal from gravitational waves seeded early in the universe, and lead to
improved constraints on cosmological parameters when combined 
with temperature measurements. The three-year \map\ observations 
\citep{page/etal:2007} showed that polarized diffuse emission 
from our Galaxy dominates the primordial 
signal over much of the sky, making accurate estimation 
of the CMB signal at large angular scales challenging. 

In this paper we describe a Bayesian framework for estimating 
the low resolution polarized CMB maps, and errors marginalized over possible 
Galactic emission. 
The goal of this approach is to determine not only the `best' 
estimate of the microwave background polarization fluctuations but to 
determine the uncertainties associated with foreground removal.
A similar technique has also been developed 
by \citet{eriksen/etal:2006,eriksen/etal:2007} for estimating 
intensity maps, and has been applied to the three-year \map\ temperature maps. 
This work complements the primary analysis of the five-year
\map\ polarization maps described in \citet{gold/etal:prep}, 
which uses template cleaning to estimate the CMB polarization maps.
The basic five year \map\ results are summarized in \citet{hinshaw/etal:prep}.
This paper is structured as follows. In \S\ref{sec:method} we 
describe the sampling method used to estimate polarized maps. 
In \S\ref{sec:sims} we apply it to 
simulated maps, and in \S\ref{sec:results} to the \map\ data. 
We conclude in \S\ref{sec:discuss}.

\section{Estimation of polarization maps}
\label{sec:method}

The large-scale polarized radiation observed by 
\map\ is the sum of at least three components: the 
primordial CMB, synchrotron emission, and thermal dust emission. Here we
briefly review the Galactic emission mechanisms, for more 
details see e.g., \citet{page/etal:2007}. 
Both synchrotron and thermal dust emission are polarized  
due to the Galactic magnetic field, 
measured to have a coherent spiral structure parallel to the Galactic plane, 
as well as a significant turbulent component 
\citep{spitzer:1998,beck:2001,vallee:2005,han:2006b}. 
The effective strength of the field is of order $\sim 10$ $\mu$G, 
thought to be split roughly 
equally between the coherent and turbulent components 
\citep{crutcher/heiles/troland:2003}.
Synchrotron emission is produced by relativistic cosmic-ray electrons 
accelerated in this magnetic field (see \citet{strong/moskalenko/ptuskin:2007} for a review of cosmic ray propagation).   
For electrons with a power law distribution of energies 
\be
N(E)\propto E^{-p},
\ee 
the frequency dependence of the emission is characterized by antenna 
temperature 
$T(\nu)\propto \nu^\beta$ with spectral index $\beta=-(p+3)/2$, with typically 
$\beta\sim-3$ \citep{rybicki/lightman:1979}. 
However, since synchrotron loss is proportional to $E^2$, 
older sources of electrons should have a lower energy 
distribution and a steeper spectral index of synchrotron emission, compared to 
regions of recently injected electrons. This leads to a synchrotron 
index that varies over the sky \citep{lawson/etal:1987,reich/reich:1988} 
and is expected to steepen away from the Galactic plane 
\citep{strong/moskalenko/ptuskin:2007}, with evidence of 
this behavior seen in the \map\ data \citep{bennett/etal:2003}.
Since the cosmic-ray electrons emit radiation almost perpendicular to the 
Galactic magnetic field in which they orbit, they can produce  
polarization fractions as high as $\sim 75\%$ \citep{rybicki/lightman:1979}, 
although integration of multiple field directions along a line of sight 
reduces this level. The fractional polarization observed at radio 
frequencies in the range 408 MHz - 2.4 GHz is further 
lowered due to Faraday rotation 
\citep{duncan/etal:1995,uyaniker/etal:1999,wolleben/etal:2006}. In the WMAP 23 GHz data, the polarization fraction is as high as 50\% on significant portions of the sky \citep{kogut/etal:2007}.

Thermal dust intensity has 
been well measured by the {\it IRAS} 
and \cobe\ missions 
and extrapolated to microwave frequencies by 
\citet{finkbeiner/davis/schlegel:1999}. 
Polarization arises since grains tend to 
align their long axes perpendicular to the Galactic magnetic 
field via, for example, the Davis-Greenstein 
mechanism \citep{davis/greenstein:1951}, and depending on their composition 
can be polarized up to a modeled 
maximum of $\sim 20\%$ parallel to 
the long axes (e.g., \citet{hildebrand/dragovan:1995,draine/fraisse:prep}). 
Observations of starlight, polarized perpendicular to the dust grains, are 
consistent with this picture \citep{heiles:2000,berdyugin/etal:2001}, 
as are the three-year \map\ observations 
\citep{page/etal:2007,kogut/etal:2007}. 
A population of smaller spinning dust grains formed of polycyclic aromatic 
hydrocarbons may also emit a significant 
amount of microwave radiation due to electric dipole 
rotational emißssion \citep{draine/lazarian:1999,draine/li:2007}. 
This question is discussed in e.g. 
\citet{hinshaw/etal:2007,dobler/finkbeiner:2007,gold/etal:prep} 
with respect to the intensity signal
observed by \map. However, these small spinning dust grains are not 
expected to be significantly polarized \citep{draine:2003}. Other
mechanisms for producing polarized emission, including 
magnetic dust \citep{draine/lazarian:1999}, 
have not been observed to be dominant.

Given these polarized Galactic components, the 
standard method used to clean the \map\ polarization 
maps involves subtracting synchrotron and dust template maps from 
the total, leaving a cleaned CMB map at the Ka, Q, and V bands 
\citep{page/etal:2007,gold/etal:prep}. The spectral indices of the templates 
are not allowed to vary spatially, which is a sufficient approximation given
the sensitivity of the observations. 
Errors are propagated by inflating the noise matrices 
to account for the uncertainties in the fitted coefficients of template maps 
\citep{page/etal:2007}.
In this alternative method we 
parameterize the emission model, and use a sampling method to 
estimate the marginal mean posterior CMB Q and U maps in low resolution pixels.

\subsection{Bayesian estimation of sky maps}
\label{subsec:bayes}

The data, $\mb d$, consist of the Q and U polarization maps 
observed by \map\  at $N_c$ frequency channels, and is a vector of 
length $2N_p \times N_c$. In this analysis we will use 
HEALPix \nside=8 with $N_p=768$ \footnote{The number of pixels is $N_p=12N^2_{\rm side}$, with $\nside=2^{\rm res}$ where res is the `resolution' 
\citep{gorski/etal:2004}.}. The joint posterior distribution for a
model $\mb m$ can be written as 
\be
p(\mb m| \mb d) \propto p(\mb d|\mb m)p(\mb m),
\ee 
with prior distribution $p(\mb m)$ and Gaussian likelihood
\be
 -2\ln p(\mb d|\mb m) = \chi^2  + c,
\ee
with 
\be
\chi^2=(\mb d- \mb m)^T \mb N^{-1}(\mb d - \mb m)
\ee
and a normalization term $c$.
Since the noise $\mb N$ is uncorrelated between channels, the likelihood can 
be written as the sum over frequency $\nu$,
\be
\chi^2=\sum_\nu [\mb d_\nu- \mb m_\nu]^T \mb N_\nu^{-1}[\mb d_\nu - \mb m_\nu],
\label{eqn:like}
\ee
where $\mb N_\nu$ is the noise covariance at each channel. In this analysis we 
use \map\ low resolution maps with inverse 
noise matrices that describe the noise outside a processing 
mask covering $\sim 6\%$ of the sky (see e.g., \citet{jarosik/etal:2007}). 
These masked pixels should be neglected in the likelihood evaluation, but to 
simplify numerical implementation we include them,
but set their inverse variance to be small (less than 0.1\% 
of the unmasked pixels' inverse variance).

We parameterize the model in antenna temperature 
as the sum of three components: 
\be
\mb m_\nu = \sum_k \bga_{k,\nu} {\mb A}_k,
\ee
with CMB ($k$=1), synchrotron emission ($k$=2), and thermal dust emission 
($k$=3).
In this analysis we will ignore possible 
polarized contributions from other components including 
spinning dust, and free-free emission. Free-free emission 
may become slightly polarized due to Thomson scattering by 
electrons in HII regions. 

The components each have amplitude vectors $\mb A_k$ of length 2$N_p$ and 
diagonal coefficient matrices $\bga_{k,\nu}$  of side 
2$N_p$ at each frequency. 
The CMB radiation is black-body, and we further assume that the Galactic 
components can be described with a spectral index that does not vary 
with frequency in the \map\ range. The coefficients are therefore given by
\ba
\bga_{1,\nu} &=& f(\nu)\mb I,\\ 
\bga_{2,\nu} &=& {\rm diag}[(\nu/\nu_K)^{\bgb_2}], \\
\bga_{3,\nu} &=& {\rm diag}[(\nu/\nu_W)^{\bgb_3}].
\ea
Here we have introduced two spectral index vectors $\bgb_k$ each of length 
2$N_p$, pivoted at 23 GHz ($\nu_K$) and 94 GHz ($\nu_W$) respectively. The function $f(\nu)$ converts thermodynamic to antenna temperature. 
We then make three further simplifying assumptions. First, that 
the spectral indices 
in Q and U are the same in a given pixel, which equates to assuming that 
the polarization angle does not change with frequency. Second, the spectrum of 
thermal dust is assumed to be fixed over the sky, with fiducial 
value $\bgb_d=1.7$, motivated by \citet{finkbeiner/davis/schlegel:1999}. 
Third, 
we define $N_{i} < N_p$ pixels within 
which $\beta_2$ takes a common value,
rather than allow it to take a unique value at each \nside=8 pixel. 
This is motivated  by our understanding of the emission process: 
even though we expect spatial variation due to the different ages of the 
electron populations, the electron diffusion rate limits how much the index 
can vary over a short range \citep{strong/moskalenko/reimer:2000,strong/moskalenko/ptuskin:2007}.  In this case we use \nside=2 HEALPix pixels ($N_i=48$). 

Our model $\mb m$ is now described by 6$N_p$ amplitude parameters $\mb A= (\mb A_1,\mb A_2,\mb A_3)^T$ and $N_i$ spectral index parameters $\bgb$. 
Our main objective is to estimate the marginalized distribution for the 
CMB amplitude vector, 
\be
p(\mb A_1|\mb d) = \int  p(\mb A,\bgb|\mb d)d \mb A_2 d \mb A_3 d\bgb,
\ee 
from which we can estimate a map and covariance matrix. 

\subsection{Sampling the distribution}
\label{subsec:sampling}

We cannot sample the joint distribution  $p(\mb A,\bgb| \mb d)$ 
directly, so we use Markov Chain Monte Carlo methods to draw samples from it. 
It can be sliced into 
two conditional distributions $p(\mb A|\bgb, \mb d)$ and 
$p(\bgb|{\bf A}, \mb d)$, so we use Gibbs sampling to draw 
alternately from each conditional distribution, constructing a Markov chain 
with the desired joint distribution as its stationary distribution. 

We briefly review Gibbs sampling for the case of one 
$A$ parameter and one $\beta$ parameter: starting from an
arbitrary point $(A_i,\beta_i)$ in the parameter space,
we draw 
\be
(A_{i+1}, \beta_{i+1}), (A_{i+2}, \beta_{i+2})... 
\ee
by first 
drawing $A_{i+1}$ from $p(A|\beta_i,d)$ 
and then drawing $\beta_{i+1}$ from $p(\beta|A_{i+1},d)$. Then 
we iterate many times. The result is a Markov chain whose stationary 
distribution is $p(A, \beta|d)$. A description of Gibbs sampling can be found
in \citet{gelfand/smith:1990}, \citet{wandelt/larson/lakshminarayanan:2004}, and 
\citet{eriksen/etal:2007}.

For the multivariate case $\mb A$ and 
$\bgb$ are now vectors, and so each vector is drawn in turn until convergence, 
producing a chain whose  stationary 
distribution is the joint posterior distribution.  
Two distinct methods are used to draw the samples from each 
conditional distribution, depending on whether the amplitude vector $\mb A$, 
or the index vector $\bgb$, is held fixed.

\subsubsection{Sampling the amplitude vector} 
For fixed $\bgb$, the conditional distribution 
$p(\mb A|\bgb, \mb d)$ is a $6N_p$-dimensional 
Gaussian, so one can draw a sample of all $6N_p$ amplitude 
parameters simultaneously. The conditional distribution is
\be
p({\bf A}|\bgb, \mb d) \propto p(\mb d|\bgb,\mb A)p(\mb A).
\ee
For a uniform prior on $\mb A$, the mean, ${\hat \mb A}$, is 
found by minimizing 
\be
\chi^2 =\sum_\nu [\mb d_\nu- \sum_k \bga_{k,\nu} {\mb A}_k]^T \mb N_\nu^{-1}[\mb d_\nu - \sum_k \bga_{k,\nu} {\mb A}_k]
\ee
with respect to $\mb A$. This gives ${\hat {\mb A}} = \mb {F}^{-1} \mb{x}$, 
which can be written in block-matrix form,
\be
\left( \begin{array}{c} \mb {\hat A}_{1} \\ \mb {\hat A}_{2} \\ \mb {\hat A}_{3} \end{array} \right) =
\left( \begin{array}{ccc} 
\mb F_{11}& \mb F_{12} & \mb F_{13} \\
\mb F_{21}& \mb F_{22} & \mb F_{23} \\
\mb F_{31}& \mb F_{32} & \mb F_{33} \end{array} \right)^{-1}
\left( \begin{array}{c} \mb x_{1} \\ \mb x_{2} \\ \mb x_{3} \end{array} \right)
\ee
with elements
\ba
\mb F_{kk'} &=& \sum_{\nu}\bga^T_{k,\nu} \mb N^{-1}_\nu \bga_{k',\nu},\\
\mb x_k   &=& \sum_{\nu}\bga^T_{k,\nu} \mb N^{-1}_{\nu} \mb d_\nu.
\ea 
Note that $\mb F$ is a $2N_p \times 2N_p$ matrix and $\mb x$ is a vector 
of length $2N_p$. 
The covariance of the conditional distribution is given by $\mb {F}^{-1}$.
In the case of diagonal noise, the mean and variance are estimated 
pixel by pixel using the same method.
Given the mean and Fisher matrix of the conditional distribution, we 
draw a Gaussian sample using the lower Cholesky decomposition of the 
Fisher matrix, $\mb F= \mb L \mb L^T$, with sample 
$\mb A_{i+1}= {\hat \mb A} + \mb L^{-1} \mb G$. The vector $\mb G$ contains 
$2N_p$ zero mean unit variance Gaussian random samples. 

For a diagonal Gaussian prior on $A_{x,k}$ of $a_k \pm \sigma_k$, 
the expressions are modified to
\ba
\tilde{\mb F}_{kk'} &=& \mb F_{kk'} + \delta_{kk'} \sigma^{-2}_{k} \mb I,\\
\tilde{\mb {x}}_k   &=& \mb x_k + \sigma^{-2}_k \mb I \mb a_k ,
\ea
with posterior mean ${\hat {\mb A}} = {\tilde \mb {F}}^{-1} {\tilde \mb{x}}$ and variance ${\tilde \mb F}^{-1}$.
In this analysis we place uniform priors on the CMB 
and synchrotron Q and U amplitudes at each pixel, but 
impose a Gaussian prior on the dust 
Stokes vector $\mb A_2=(\mb Q_2,\mb U_2)^T$ of 
$[\mb Q_2,\mb U_2]=0\pm0.2~\mb I_d$, using the 
dust map ${\bf I}_d$ at 94 GHz from model 8 of 
\citet{finkbeiner/davis/schlegel:1999}, 
hereafter FDS, as a tracer of the intensity.  The width of the prior, 
corresponding to a polarization fraction 20\%, 
is motivated by \citet{draine/fraisse:prep}, who predict the 
maximum polarization of dust grains to be about 15\%.

Drawing this new amplitude vector is computationally demanding, and 
drives us to work with low resolution maps. Our 
goal is to determine the polarized CMB signal at large 
angular scale, so this does not limit the analysis.

\subsubsection{Sampling the index vector}
 
For fixed A, we sample from the conditional distribution 
\be
p(\bgb|\mb A, \mb d) \propto p(\mb d|\bgb,\mb A)p(\bgb),
\ee
with prior probability $p(\bgb)$. An analytic sample cannot be drawn from 
this distribution because the spectral indices are non-linear parameters. 
However, for a 
small number of parameters it is feasible to draw samples 
using the Metropolis-Hastings algorithm. 
This algorithm has been described extensively in the cosmological parameter
estimation literature (e.g., 
\citet{knox/christensen/skordis:2001,lewis/bridle:2002,dunkley/etal:2005}).
The sampling goes as follows. For each index parameter 
in turn, a trial step $\beta_T$ 
is drawn using a Gaussian proposal distribution of width $\sigma_T$ 
centered on the current $\bgb$ vector.
Next, the current and trial $\bgb$ vectors are used to construct 
model vectors at each frequency, $\mb m_\nu=\sum_k \bga_{k,\nu} \mb A_k$. 
The current and trial posterior are then computed using
\be
-2\ln p(\bgb|\mb A,\mb d)= \chi^2  -2\ln p(\bgb),
\ee  
with $\chi^2$ given in Eqn \ref{eqn:like}. 
The ratio of the trial to current posterior, $r$, is used to determine 
whether to move to the trial position (with probability $r$), or to stay 
at the original position (with probability $1-r$).
This use of the Metropolis algorithm to draw a subset of parameters 
is commonly known as Metropolis-within-Gibbs 
(e.g. \citet{geweke/tanizaki:2001}), 
and has been used in astronomy to estimate Cepheid distances 
\citep{barnes/etal:2003b}. Other approaches to sampling spectral 
index parameters have been considered in e.g., \citet{eriksen/etal:2007}.

In regions of low signal-to-noise, it is necessary to impose a prior on the 
synchrotron spectral index, otherwise it is unconstrained and could take
the `flat' index of the CMB component, opening up large degeneracies. 
We choose a Gaussian prior of $-3.0\pm0.3$, motivated by understanding of the
synchrotron emission \citep{rybicki/lightman:1979} and allowing for 
variations of the size observed in the synchrotron intensity 
(e.g., \citet{bennett/etal:2003}). This is combined with a 
uniform prior on the CMB and synchrotron amplitudes, and a Gaussian 
prior $0\pm0.2I_d$ on the dust Q and U amplitudes.
This parameterization and choice of prior does not guarantee that the 
marginalized means for the $A$ and $\beta$ parameters 
will be unbiased estimators, as discussed in e.g. \citet{eriksen/etal:2007}.
In the limit of low signal-to-noise, there 
is a larger volume of models with a flatter (i.e. $\beta$ tending to 0) 
synchrotron spectrum, allowing large CMB and synchrotron amplitudes 
of opposite sign. One approach is to modify the prior on the spectral indices 
to include an additional 
`phase-space' factor. We discuss this further in \S\ref{sec:sims}.

\subsection{Processing the sampled distribution}
\label{subsec:process}

We form maps of each component from the mean of the marginalized distribution,
\be
\langle \mb A_k \rangle =\int p(\mb A_k |\mb d)\mb A_k d\mb A_k = \frac{1}{n_{\rm G}}\sum_{i=1}^{n_{\rm G}} \mb A_k^i,
\ee 
where the sum is over all $n_{\rm G}$ elements in the chain, and 
$\mb A_k^i$ is the $i$th chain element of the $k$th component map. 
The covariance matrices for $\mb A_k$, including off-diagonal terms, 
are estimated using the same method, summing over
the chain components. As an example, the covariance $C_{xy,k}$ between 
pixels $x$ and $y$ for component $k$ is computed using
\ba
C_{xy,k} &=& \langle A_{x,k} A_{y,k}\rangle - \langle A_{x,k} \rangle \langle A_{y,k} \rangle\\
  &=& \frac{1}{n_{\rm G}}\sum_{i=1}^{n_{\rm G}} (A^{i}_{x,k}- \langle A_{x,k} \rangle)(A^{j}_{y,k}-\langle A_{y,k} \rangle).
\ea
For the synchrotron and dust components, we compute only the diagonal 
elements of the covariance matrices.

\subsubsection{Power in the E-mode}

To quantify the polarization anisotropy present in the Q and U maps, 
we use the coordinate-independent 
scalar and pseudo-scalar E and B modes commonly used in cosmological analysis 
\citep{seljak:1997,kamionkowski/kosowsky/stebbins:1997}.  
Both polarization modes probe the evolution of 
the decoupling and reionization epochs and are generated by 
Thomson scattering of a quadrupolar radiation 
pattern by free electrons. 
The anisotropy is quantified using the $C_\ell^{TE}$,  $C_\ell^{EE}$, 
$C_\ell^{BB}$ power spectra, where
\be
C_\ell^{XY}= \langle a_{lm}^Xa_{lm}^{Y*} \rangle.
\ee
The spin-2 decomposition of the polarization maps, $a^{E,B}_{lm}$, 
is related to the Q and U maps by 
\begin{equation}
[Q\pm iU](\hat{x}) = \sum_{\ell>1}
\sum_{m=-\ell}^{\ell}{{}_{\mp2}a_{\ell m} \,{}_{\mp2}Y_{\ell m}(\hat{x})}
\end{equation}
where $_{\pm2}a_{lm}=a_{lm}^E\pm i a_{lm}^B$ \citep{zaldarriaga/seljak:1997}. 

The first stars reionize the universe at redshift 
$z_r$, producing a 
signal in the E-mode power spectrum proportional to $\tau^2$, 
where $\tau$ is the optical depth to reionization. 
In this analysis we use the approximation to the optical depth 
used in \citet{page/etal:2007}, estimated by varying only $\tau$ and the 
power spectrum amplitude, such that the temperature anisotropy 
power at $\ell=220$ is held constant. Other cosmological parameters are fixed 
to fiducial values 
and the exact likelihood is computed as a function of $\tau$.
The likelihood, $p(C_\ell|\mb d)$, given in Appendix D of 
\citet{page/etal:2007}, is evaluated using the 
marginal posterior mean Q/U maps and their covariance matrix, which are 
processed as described in \citet{page/etal:2007} to account 
for the P06 Galactic mask.

\subsection{Testing convergence}

The convergence of the chain is determined by applying 
the spectral convergence test described in \citet{dunkley/etal:2005} 
to the spectral index
parameters and a random subset of the amplitude parameters.
This tests for convergence of the mean, but convergence of the 
covariance matrix is also required for estimating 
the power in the mapx at large 
scales. We check this  
by applying a jackknife test to the derived optical depth parameter $\tau$. 
For a chain that samples only the amplitude vectors, and has diagonal 
noise, 1000 iterations are typically sufficient. With the 
Metropolis sampling step included, about 10,000 iterations are required, and
when off-diagonal noise is included, typically 50,000 iterations are necessary.

\section{Simulated polarization maps}
\label{sec:sims}

We simulate all-sky signal and noise maps at HEALPix \nside=16, with 
$N_p=3072$ pixels, for 
the Stokes Q and U parameters, at the five \map\ frequencies (22.8, 33, 40.7, 60.8, and 93.5 GHz). In the notation of \S\ref{sec:method}, the data are 
modeled as 
\be
\mb s_\nu =f(\nu)\mb A_1 + (\nu/\nu_K)^{\beta_s}{\mb A}_2 + 
(\nu/\nu_W)^{\beta_d}{\mb A}_3,
\ee
where $\mb s_\nu$ is a vector of length 2$N_p$ containing the total 
Q and U signal in antenna temperature at frequency $\nu$.
The CMB signal map, $\mb A_1$, is generated for a fiducial 
cosmological model by drawing multipoles $a_{lm}$ from the 
theoretical power spectrum $C_\ell$, and transforming to map space, using the 
`synfast' routine in HEALPix. We choose a model with $\tau=0.1$, 
with other cosmological parameters given by the best-fit 
three-year \map\ $\Lambda$CDM model \citep{spergel/etal:2007}. 
For the synchrotron emission, with amplitude map $\mb A_2$ defined at 23~GHz, 
we use the three-year \map\ Q and U  low-resolution K-band maps
\citep{page/etal:2007}. The frequency dependence is assumed to be a power 
law with $\beta=-3.0$ over the full sky. For the thermal dust map, 
$\mb A_3$, defined at W band (94 GHz),  we construct maps 
that are 5\% polarized, 
using the dust intensity template $\mb I_d$ from IRAS, extrapolated to 94~GHz 
using Model 8 of \citet{finkbeiner/davis/schlegel:1999} 
and degraded to \nside=16. We form 
\ba
\mb Q_3 &=& 0.05 \mb I_d\cos(2\gamma)\\
\mb U_3 &=& 0.05 \mb I_d\sin(2\gamma),
\ea 
where the polarization angles $\gamma$ are computed from the starlight 
polarization template described in \citet{page/etal:2007}, using
observations from \citet{heiles:2000,berdyugin/etal:2001,berdyugin/piirola/teerikorpi:2004}. 
We assume a power law index with $\beta_d=1.7$ 
over the whole sky. This is a simplified model of the true 
sky, as we would expect the emission processes to result in some 
spatial and frequency dependence of the spectral indices. A realistic Galactic field model would also lead to a spatially dependent suppression 
of the dust polarization fraction 
\citep{page/etal:2007,miville-deschenes/etal:prep}.

The simulated maps are formed at each frequency $\nu$ 
using $\mb d_\nu = \mb s_\nu +\mb n_\nu$, 
where $\mb n_\nu$ is a realization of the \map\ noise. We consider both 
diagonal noise realizations, and full noise realizations including 
pixel-pixel and Q-U correlations, drawn from the 
\map\ \nside=16 noise matrices. These correspond to maps 
co-added by year and DA, as described in \citet{jarosik/etal:2007, hinshaw/etal:prep}. 
The low resolution inverse noise matrices are not defined inside the 
processing mask, so we set the noise in these pixels to be large. 
Realizations are computed at each channel 
using SVD decompositions of the \map\ inverse noise matrices
$\mb N^{-1}_\nu$, in antenna temperature.
We then create \nside=8 simulations, with 768 pixels, by degrading the 
\nside=16 simulated maps and $\mb N^{-1}$ inverse noise matrices 
using inverse weighting.

We perform the Gibbs sampling using the four 
\map\ frequency bands K, Ka, Q, V (hereafter KKaQV).  We do not include
W band as standard, due to concerns about
potential systematic effects \citep{hinshaw/etal:prep}.
For the diagonal noise case this 
requires of order 10,000 iterations for convergence. 
The chains are processed as
described in \S\ref{subsec:process} to obtain marginal posterior mean maps 
and errors for $\mb A_1$ (containing Q and U for CMB), $\mb A_2$ (synchrotron), $\mb A_3$ (thermal dust) and $\bgb$ (synchrotron spectral index). 

\begin{figure}[t]
  \epsscale{1.2}
  \plotone{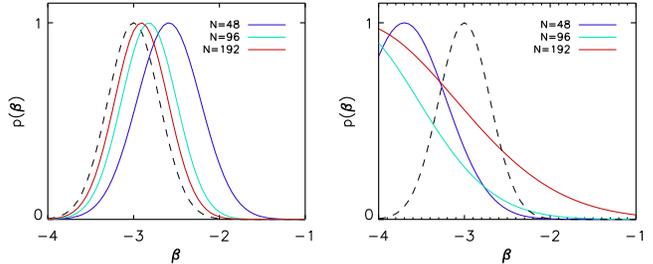}
 \caption{Left: For maps with $N_p=768$ pixels and $N_i$ (marked 
$N$ in figure) synchrotron 
spectral indices with Gaussian priors of $\beta=-3.0\pm0.3$ (dashed curve), the 
recovered index estimates are larger than the prior (solid curves) 
for simulations with no signal. The estimated index mean 
increases with the number of pixels sharing the same 
index. Right: The spectral index priors that are imposed (colored curves), in addition to $\beta=-3.0\pm0.3$, to account for the volume effect. 
 \label{fig:beta_bias} }
\end{figure}

\begin{figure*}[t]
\center{
\epsscale{1.1}
\plotone{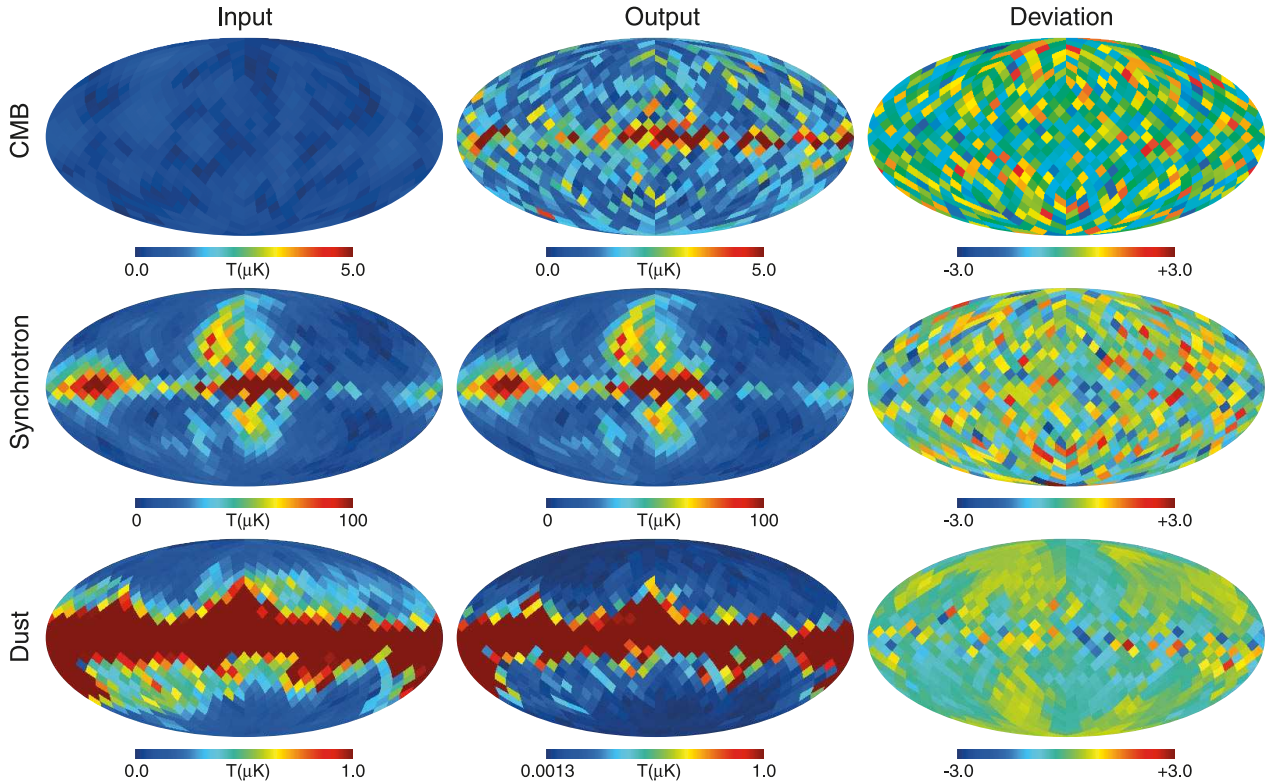}
}
\caption{Comparison of input (left) and output (middle) 
component maps for a simulation with \nside=8. 
The polarization amplitude $P=\sqrt{Q^2+U^2}$ is 
shown for the CMB (top), synchrotron at 23 GHz (middle), 
and dust at 94 GHz (bottom). 
The right panel shows the 
difference in standard deviations per pixel for the Q Stokes 
parameter.
The dust component has a Gaussian 
prior on the dust Stokes parameters of $[Q,U](n)= 0 \pm 0.2 I_d(n)$, where $I_d(n)$ is the FDS dust intensity (see text), which reduces the deviation per pixel and leads to the structure in the difference map.
\label{fig:r4_sim}}
\end{figure*}

\subsection{Spectral index prior}
\label{subsec:phase_prior}

For the simulated maps with $\beta=-3.0$ and a Gaussian prior on the indices  
of $-3.0\pm0.3$, we find 
the recovered marginal posteriors $p(\beta|d)$ to be biased estimators.
To explore this further, we sample the joint distribution for a noise-only 
simulation, drawn from the \map\ five-year noise maps. 
The resulting marginalized distributions 
for the $N_i=48$ spectral indices are close to Gaussian, but centered 
on $\beta=-2.6$. The Gaussian distribution best fitting the samples 
is shown in the left panel of Figure \ref{fig:beta_bias}. When the number of 
spectral index pixels is increased to 96, and 192, the recovered distribution
tends toward the input value of $\beta=-3.0$. This effect arises for 
our Galactic emission model, as there is a larger volume of phase 
space for shallow indices far from the prior central value. This is not 
a significant effect for individual pixels, but the phase-space 
volume relative to 
that of the prior central value increases for more pixels sharing a common 
index parameter. 

In the case of a single pixel, a common approach is to 
adopt the Jeffreys' prior \citep{jeffreys:1961}, a 
non-informative prior distribution that is proportional to the 
square root of the Fisher information, as described for example 
in \citet{eriksen/etal:2007}. Imposing the Jeffreys' prior 
on the spectral index parameters in this model 
would give less weight to steep indices 
than shallow ones, accounting for their smaller effect
on the likelihood. However, this would not overcome the phase space effect that 
arises from the increase in volume of models with shallow indices.

While there may be alternative parameterizations that avoid
this problem, we choose instead to modify the spectral index prior. 
We use $p(\beta) = p_s(\beta)/L_0(\beta)$, 
where $p_s$ is the Gaussian prior $-3.0\pm0.3$, and $1/L_0$ is an additional 
`phase-space' factor. We estimate this factor 
by evaluating the marginalized 
posterior, $p(\beta|d)$, for a noise-only simulation with prior $p_s(\beta)$, 
and defining $L_0(\beta) = p(\beta|d)/p_s(\beta)$. The assumption is that $L_0$ 
is an approximate description of the available phase-space volume. 
The factors, $1/L_0$, are modeled as Gaussian 
distributions and are shown in the right panel of 
Figure \ref{fig:beta_bias} for $N_i$=48, 92 and 192, with $\beta =-3.7\pm0.5$ 
for the $N_i=48$ adopted in this 
analysis. These distributions 
are not exact, particularly at $\beta <-4$, but we check 
that the approximation is sufficient by 
resampling the distributions with the new 
prior, finding the recovered mean to be $\beta=-3.0$. We repeat the test for 
a Gaussian prior of $-2.7\pm0.3$, and further test that this prior 
only weakly depends on the inclusion of W band. 

This is not the optimal solution, as the prior is 
informative in the presence of a signal and could therefore 
lead to parameter bias, in particular in the weak signal regime.
We test simulations for two different signal maps 
with synchrotron index 
$-3.0$ and $-2.7$, using Gaussian priors $p_s$ of $-3.0\pm0.3$ and $-2.7\pm0.3$ 
combined with the phase-space factor. 
As expected, when the prior $p_s$  matches the signal, 
the estimated index values are all consistent with the input signal. When it 
does not match the signal, an unbiased marginal index value is estimated
in the highest signal-to-noise pixels, and the low signal-to-noise pixels tend 
to the central value of $p_s$. 
This is the behavior we expect, but does not confirm that 
our final estimated parameters will be unbiased. 
However, since we have good astrophysical motivation to require 
that the synchrotron index is close to $\beta=-3$, we do not expect 
this to be a significant effect. We address this issue in this analysis 
by checking that our main
conclusions are insensitive to the detailed choice of the spectral index 
prior, and defer an investigation of a more optimal parameterization 
to a future analysis.

\subsection{Simulations with diagonal noise}

We now apply the method to simulations with signal and diagonal noise 
properties. Figure \ref{fig:r4_sim} shows the input and output 
polarization amplitude maps for the CMB, synchrotron, and dust components, 
derived from the Q and U maps, $P=\sqrt{Q^2+U^2}$, for the fiducial simulation
with synchrotron index $\beta=-3.0$. The third column 
shows the difference in standard deviations per pixel for the Q Stokes 
parameter, $\delta = (Q_{\rm in}-Q_{\rm out})/\sigma_Q$.
For the CMB maps the absolute difference 
between input and output maps is greatest in the 
Galactic plane where the foreground signal is high. However, since 
these effects are captured in the errors in the map, 
the deviation maps do not have a strong 
spatial dependence. 
The $\chi^2$ for the map, $(\mb A_1^{\rm in}-\mb A_1^{\rm out})^T \mb C_1^{-1}(\mb A_1^{\rm in}-\mb A_1^{\rm out})$, is 1270 for 1536 pixels.
This gives $\chi^2/{\rm pixel} <1$, due to the prior on the 
dust amplitude. When the prior on the dust amplitude is removed, the 
goodness of fit of the recovered CMB maps is $\chi^2=1497$ ($\chi^2/{\rm pixel}=0.98$), but the errors are significantly inflated.
The synchrotron maps are recovered with 
$\chi^2/{\rm pixel}=1.04$ for the Q and U maps.
The dust maps are recovered with $\chi^2/{\rm pixel}=0.24$, with 
structure apparent in the deviation map. This is due to 
the prior: far from the plane the 
signal-to-noise is low and the dust tends to the prior central value of 
zero. Removing the prior on the dust amplitude, the goodness of fit of
the recovered dust maps is $\chi^2/{\rm pixel}=0.97$, close to 1 as expected. 
The total $\chi^2$ of the estimated model, $\sum_\nu [\mb d_\nu- \mb m_\nu]^T \mb N_\nu^{-1}[\mb d_\nu - \mb m_\nu]$, is $2.08N_p$ without the dust prior, and $3.51N_p$ with the dust prior. This is consistent with the number of degrees of freedom ($2N_p- N_i$ in the absence of priors).

\begin{figure}[t]
  \epsscale{1.1}
\center{
  \plotone{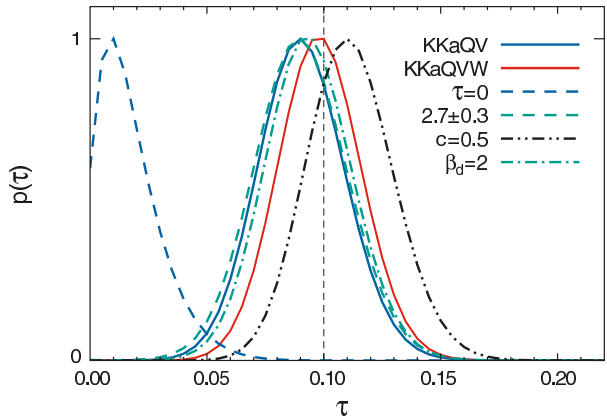}
}
  \caption{The distribution of the optical depth to reionization $\tau$ for 
simulations with five-year \map\ noise levels. The optical depth is 
recovered for an individual simulation with $\tau=0.1$ (black dashed line), for 
KKaQV and KKaQVW, and a $\tau=0$ simulation is consistent with zero power. 
Changing the index priors to $\beta_d=2$, or $\beta_s=-2.7\pm0.3$ have 
negligible effects on the recovered CMB power. Incorrectly modeling 
the synchrotron as a power-law for 
a simulation with an index that increases by $c=0.5$ between 23 and 94 GHz 
increases $\tau$ by $\sim$1$\sigma$.
 \label{fig:tau_sim} }
\end{figure}

The optical depth computed from this simulation is 
$\tau=0.091\pm0.019$ outside the P06 Galactic mask, with distribution shown 
in Figure \ref{fig:tau_sim}. 
We test for bias by generating 
10 further diagonal noise and signal realizations of the model 
with $\tau=0.1$, and find ensemble average of $\tau=0.098$, consistent with
the input but limited by small number statistics. We also test a 
simulation for 
KKaQV with no polarized CMB component. The recovered optical depth 
to be consistent with zero, shown in Figure \ref{fig:tau_sim}.
Adding W band (KKaQVW) we find $\tau=0.098\pm0.017$. 

In this ideal case the simulation 
matches the parameterized model, but in a realistic 
scenario the model will not perfectly describe the sky. For a Gaussian 
prior on the spectral index of $\beta_s -2.7\pm0.3$, 
we find a negligible effect on $\tau$, with
$\tau=0.094\pm0.020$. A larger increase of $\sim$1$\sigma$ in $\tau$, 
to $0.111\pm0.019$, is seen when the 
spectral index of the simulation shallows from -3.0 to -2.5  
between K and W band, but is modeled by a pure power law. In this case
the model does not remove enough synchrotron at high frequencies, leaving
excess power in the CMB.
For the thermal dust component we fix $\bgb_d=1.7$ for all pixels in the 
fiducial case. Changing this 
to $\bgb_d=2$ has a negligible effect on our results.
Removing the dust prior altogether opens up long degeneracies between 
the dust and the CMB amplitudes, highlighting 
the importance of the dust intensity map to limit 
the polarized dust contribution. Modest changes of the dust polarization
fraction prior to 15\% or 25\% have little 
effect on the estimated CMB signal.

\subsection{Simulation with full noise matrix}
\begin{figure}[t]
    \epsscale{1.1}
  \plotone{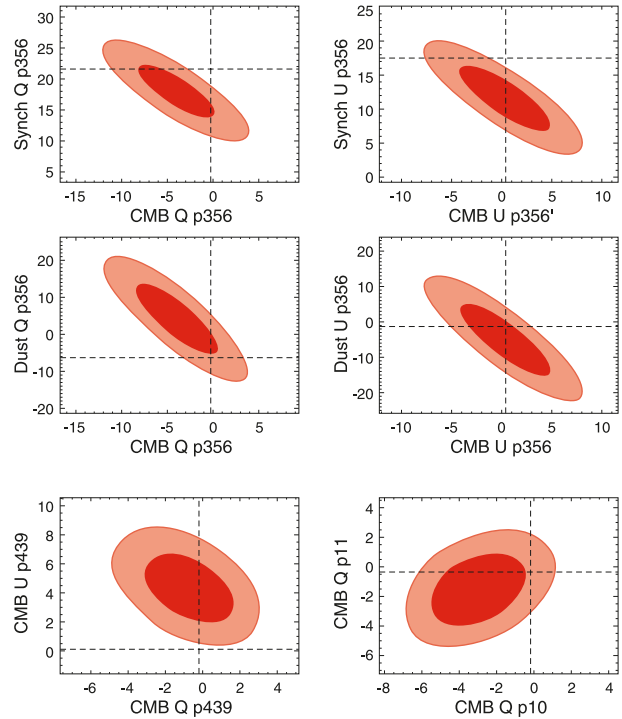}
  \caption{Marginalized 68\% and 95\% confidence levels for a subset of the 
$A$ and $\beta$ parameters for simulated maps (inputs shown dashed). 
The top and middle panels show the 
correlation between the CMB, synchrotron and dust Q and U amplitudes (in $\mu K$) 
for an arbitrary 
single pixel (pixel 356 out of 768 using HEALPix nested ordering). 
The bottom panels show the correlation between Q and U for a single 
pixel (p439), and between two adjacent pixels (p10 and p11). By marginalizing, rather than finding the maximum likelihood, the error on the CMB amplitude is inflated to account for foreground uncertainty. 
  \label{fig:twod_sim} }
\end{figure}

When the full off-diagonal noise matrix is included we 
find of order 50,000 iterations are 
necessary for convergence.
The $\chi^2$ for the map computed using the full
covariance matrix, 
$(\mb A_1^{\rm in}-\mb A_1^{\rm out})^T \mb C_1^{-1}(\mb A_1^{\rm in}-\mb A_1^{\rm out})=1220$  
for 1536 pixels, consistent with the diagonal noise case ($\chi^2=1270$). 
The recovered spectral index distributions are consistent with the 
simulated values, and the estimated optical depth of a single simulation 
with input $\tau=0.1$ is $0.110\pm0.020$. Testing a large set of 
simulations with full inverse noise matrices is beyond the scope of this 
analysis. 
To demonstrate the effect on the estimated CMB maps of 
marginalizing, Figure \ref{fig:twod_sim} shows two-dimensional marginalized 
distributions for a subset of parameters, for a single pixel and 
between pixels. The top and middle rows show the 
correlation between the CMB, synchrotron, and dust Q and U components within 
a single pixel. The one-dimensional 
error on the marginalized CMB Q and U amplitudes is larger than the error obtained 
if the foreground amplitudes are fixed at 
their maximum likelihood amplitudes. The 
bottom row shows correlations between Q and U components within a pixel, 
and between adjacent pixels. If the inter-pixel correlations are ignored, 
the maps are recovered with incorrect noise properties.

\begin{figure*}[t]
  \epsscale{1.0}
  \plotone{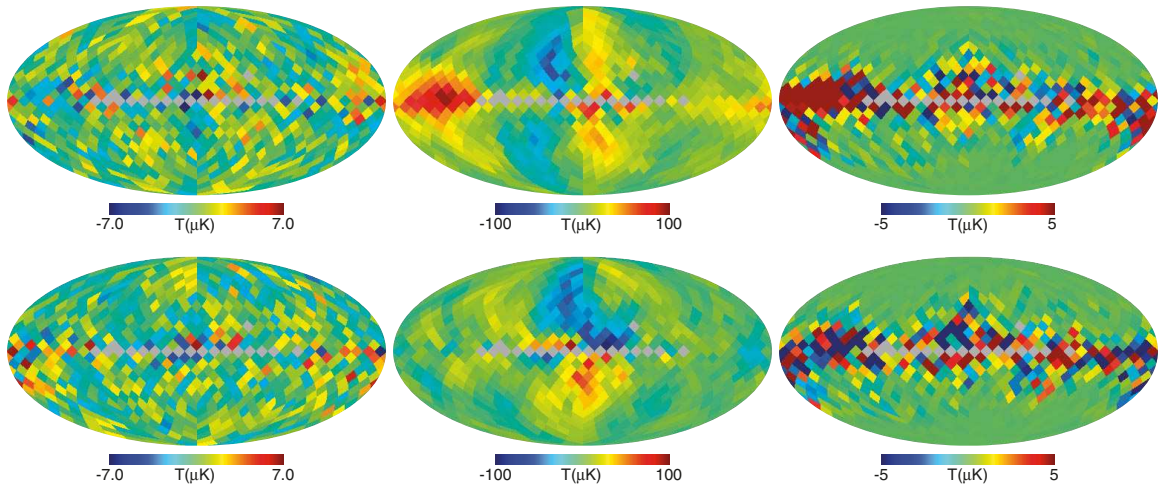}
  \caption{Low resolution polarized Q (top) and U (bottom) maps 
of the CMB, synchrotron, and dust emission, estimated from the five-year 
K, Ka, Q, and V band maps using Gibbs sampling. 
Pixels inside the processing mask are grey.
The CMB maps (left panels, thermodynamic temperature) 
do not show significant Galactic contamination 
in the plane. The synchrotron amplitudes (center, antenna temperature), are 
defined at K-band (22.8 GHz), and are 
consistent with the total K-band maps, 
with high Q and U emission from the North Polar Spur, 
and high Q emission in the Galactic plane at longitude $110\ltsim l \ltsim 170$. The dust amplitudes (right, antenna temperature) are defined at W-band 
(93.5 GHz), and have a Gaussian prior on Q and U of $0\pm0.2I_d$
where $I_d$ is the dust intensity.
\label{fig:pol_map}}
\end{figure*}

\begin{figure*}[t]
 \epsscale{1.0}
 \plotone{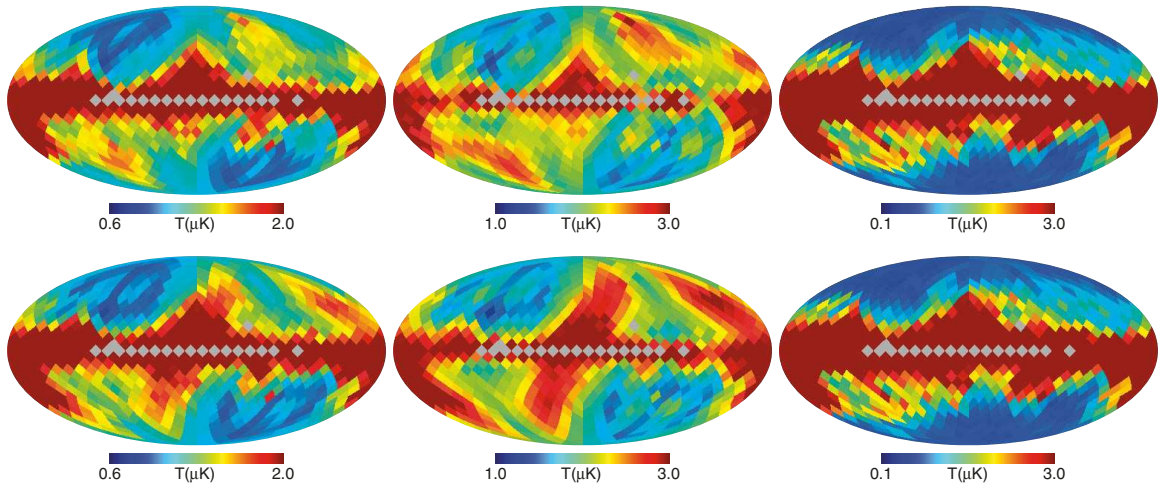}
\caption{Estimated 1$\sigma$ errors on the low resolution maps of the CMB 
(left), synchrotron (center), and dust (right) Q and U components, 
as shown in Figure \ref{fig:pol_map}. The 
CMB errors are more fully described by a covariance matrix, including 
pixel-pixel correlation and Q/U correlation, so the maps can be used for
cosmological analysis. 
The errors on the dust 
maps (right) are dominated by the prior that limits the 
dust polarization fraction to 20\%.
The middle panels clearly show the two sources of uncertainty in 
our CMB polarization maps: detector noise in the ecliptic plane 
(which traces a sideways S in the map) and foreground removal 
uncertainties in the galactic plane.
\label{fig:pol_err}}
\end{figure*}

\section{Results from \map\ data}
\label{sec:results}

We apply the sampling method to the five-year \map\ data, using 
low resolution coadded \nside=16 maps and inverse noise matrices as the 
inputs, and then degrading to \nside=8 as for the 
simulations. Figures \ref{fig:pol_map} and \ref{fig:pol_err} 
show maps of the mean values and 1$\sigma$ errors for the 
marginalized CMB, synchrotron, and dust amplitudes. 

\subsection{CMB polarization}
We first consider the CMB results. The noise patterns for both Q and U 
in Figure \ref{fig:pol_err} are consistent with what 
we expect: in regions of low Galactic emission, the errors are 
dominated by instrumental noise. As the Galactic plane is approached, 
errors from foreground uncertainty begin to dominate, in particular where the 
dust contribution is most uncertain.
This is a real advantage of the method: rather than imposing masks, the 
method inflates the errors where the foregrounds are brightest. 
As opposed to template cleaning, which produces CMB maps at 
each frequency observed, this method recovers a single Q and U 
polarization map, and so has higher signal-to-noise than any of the 
individual template-cleaned maps.  There is some indication of structure 
in the CMB signal, particularly in the Q Stokes map at the Galactic 
anti-center, and in the U map in the region of the North Polar Spur. This is 
consistent with noise. 
Outside the P06 mask the maps are morphologically similar to 
the template-cleaned CMB maps co-added over Ka, Q, and V bands. 
The correlation coefficients for the pixels within the CMB maps have a maximum value of $\sim 40\%$, with rms correlation of $\sim 2\%$.

\begin{figure}[t]
\epsscale{1.1}
  \plotone{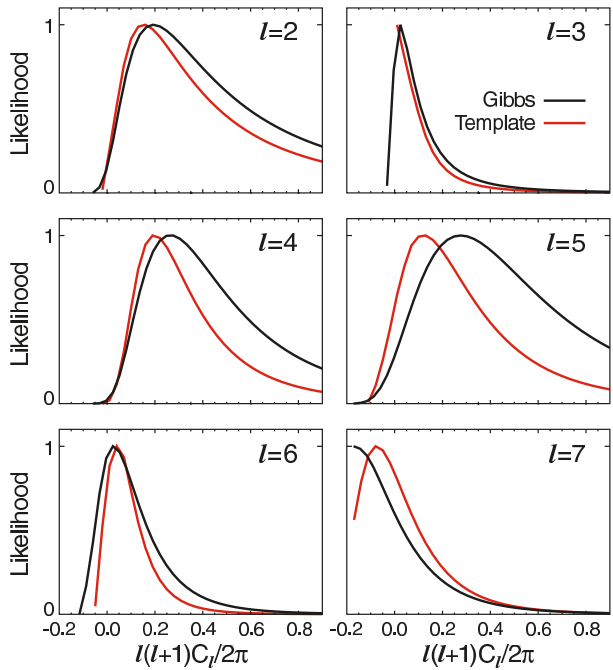}
  \caption{Conditional likelihoods for the CMB EE multipole moments 
estimated from the polarization maps described in this analysis (black), 
compared to the template cleaned maps described in \citet{gold/etal:prep} 
(red).
They are computed as in \citet{nolta/etal:prep} 
by fixing all other $C_\ell$ values at the 
fiducial $\Lambda$CDM values. There is agreement between the two methods.  
  \label{fig:ee_spec} }
\end{figure}

We compare the power at each multipole, outside the P06 mask, to 
the template-cleaned case 
from the main \map\ analysis \citep{gold/etal:prep} in Figure \ref{fig:ee_spec}. Using the method 
described in \citet{nolta/etal:prep}, at each multipole 
the conditional likelihoods is computed 
as a function of $C_\ell^{EE}$, with all other multipoles held fixed, 
for $2<\ell<7$. 
The results are consistent, although this analysis finds more power 
at $\ell=4$ and $5$. Computing the likelihood as a function of $\tau$ we 
find $\tau=0.090\pm0.019$ for the Gibbs-sampled maps outside 
the P06 mask, which is 
consistent with the results obtained through template cleaning, 
which give $\tau=0.086\pm0.016$ for the KaQV data combination. 
Obtained using a different 
methodology and accounting for foreground marginalization, 
this adds confidence in the detection of the CMB E-mode
polarization signal. 
The spatially varying
synchrotron index appears not to cause a significant difference, 
as we obtain a similar mean when the synchrotron spectral coefficients are fixed at the best-fit values found in the template cleaning. This also confirms 
that the choice of the spectral index prior is not affecting the estimated CMB 
power in the maps.
We find similar limits on the optical depth 
when W band is included (KKaQVW), with $\tau=0.085\pm0.017$. 
However, we choose not to use W band in the standard analysis, as 
discussed earlier.
The Gaussian prior on the dust Q and U parameters 
does affect the CMB signal constraint: removing it 
significantly weakens the limit on the large-scale 
power, but changing it to e.g., 15\% or 25\% has little effect, 
consistent with tests on simulations. 
A significantly tighter prior is not physically motivated, and 
could lead to bias in the recovered signal.

\subsection{Synchrotron polarization}
\label{subsec:synch_dust}

\begin{figure}[t]
\begin{center}
\epsscale{0.8}
\plotone{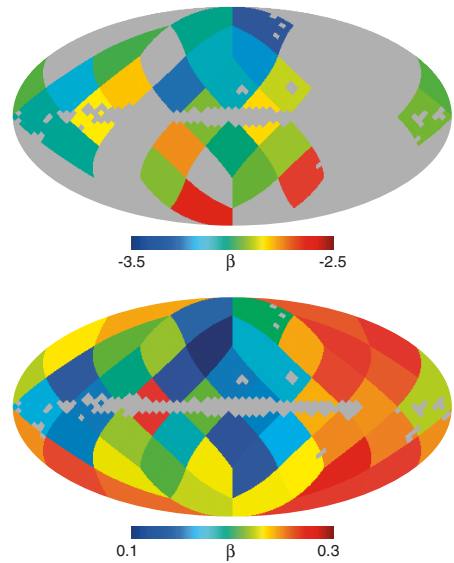}
\caption{
The estimated synchrotron spectral index (top), and 
1$\sigma$ errors (bottom), estimated 
in 48 HEALPix \nside=2 pixels. A Gaussian 
prior of $\beta_s=-3.0\pm0.3$  is imposed in each pixel.
In regions of low signal-to-noise (near the ecliptic poles), the prior 
drives the spectral index estimate, so we mask the index for pixels with 
$\sigma(\beta)>0.25$. 
The mean index averaged over the unmasked pixels 
is $-3.03\pm0.04$ for prior $-3.0\pm0.3$.
\label{fig:pol_index}}
\end{center}
\end{figure}

\begin{figure}[t]
  \epsscale{1.1}
  \plotone{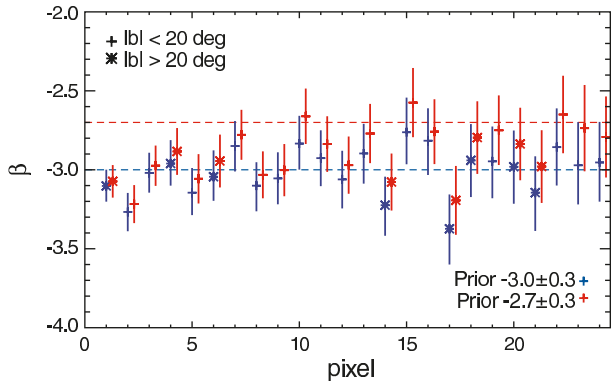}
  \caption{Marginalized mean and 1$\sigma$ error bars 
for the synchrotron spectral index parameters with highest signal-to-noise 
for the five-year \map\ data, 
for Gaussian priors of $-3.0\pm0.3$ and $-2.7\pm0.3$. The pixels are ordered with increasing 1$\sigma$ errors, with the highest signal-to-noise pixels 
on the left. The prior has only a small effect on the estimated indices.
  \label{fig:beta_wmap} }
\end{figure}

The synchrotron maps shown in Figure \ref{fig:pol_map} are 
similar to the total K-band maps \citep{hinshaw/etal:prep}. The difference
between the estimated synchrotron amplitude, and the K-band amplitude, 
is $< 5~\mu K$ outside the P06 mask, and $<8~\mu K$ in the Galactic plane. 
As observed in the three-year \map\ data \citep{page/etal:2007}, the signal 
is dominated by emission from the North Polar Spur, marked on the 
microwave sky map in \citet{hinshaw/etal:2007}, as well as what is often known 
as the `Fan region' (e.g., \citet{wolleben/etal:2006}), 
centered on Galactic coordinates  $l \sim 140$, $b \sim 5$. The 
synchrotron emission dominates the signal at low 
frequencies, and so the uncertainty in the synchrotron Q and U maps, 
shown in Figure \ref{fig:pol_err}, is dominated by 
instrument noise, with only a small contribution from 
marginalization in the Galactic plane.
Figure \ref{fig:pol_index} 
shows the mean synchrotron spectral index estimated in the
48 pixels, together with 1$\sigma$ errors.  In 
regions of low synchrotron signal-to-noise the index is driven by the 
prior, so we mask pixels with  $\sigma(\beta)>0.25$. 
There is a preference in the 
North Polar Spur and Fan region for an index of $\sim -3$: the 
index averaged over regions with $\sigma(\beta)<0.25$ is 
$-3.03\pm0.04$. The estimated indices of these  
pixels are also shown in Figure \ref{fig:beta_wmap}, ordered by 
increasing errors to highlight 
the high signal-to-noise pixels on the left. 
Cutting the sky into high and low latitude at $b=20$~degrees, the
low latitude weighted mean, $-3.00\pm0.04$, is 
consistent with the high latitude $-3.08\pm0.06$.
This contrasts with the more significant steepening of the index 
with latitude observed in \citet{kogut/etal:2007} in the 
analysis of the three-year \map\ polarization data, although here we 
exclude part of the plane. These results
coming from alternative analyses will be more easily assessed with 
higher signal-to-noise data. We check that the latitude dependence 
is unaffected when the phase-space factor in the spectral index
prior, a Gaussian $-3.7\pm0.5$, is modified to have width $1.0$ 
or mean $-4.0$.

An extrapolation of the Haslam 408 MHz synchrotron intensity maps 
to \map\ frequencies suggests that a shallower index is preferred in 
the plane, compared to the polarization, in the absence of anomalous dust. 
However, our estimated polarization index 
is consistent with the synchrotron 
intensity maps derived by the MEM method \citep{gold/etal:prep}. Degrading the
synchrotron component map in each band to \nside=2, and 
fitting a power law to the data in K, Ka, Q, and V band for each pixel, 
the spectral index values are best-fit by a power-law with 
index $\beta= -3.1$ for the pixels at $b=0$. 
 
We check the estimate of the spectral indices by re-running the analysis 
with a Gaussian synchrotron index prior, $p_s(\beta)$, 
of $-2.7\pm0.3$ in addition to the phase-space factor. 
The estimated indices are 
compared pixel by pixel in Figure \ref{fig:beta_wmap}. The prior 
has only a small effect on the high signal-to-noise index estimates.  
The $\sigma(\beta)<0.25$ pixel average in this case is $-2.94\pm0.04$.
The estimated CMB power is little affected by this change in prior, 
with $\tau=0.092\pm0.019$. 
 
\subsection{Dust polarization}

The dust polarization map has a low signal-to-noise ratio, particularly 
far from the plane, as we only fit data in the K-V bands. In these
regions the prior dominates the estimate of the dust amplitude, making
it hard to draw conclusions about the dust component. 
The error in Q and U is driven by the prior on the 
polarization amplitudes and so is morphologically identical to the FDS dust 
intensity map. This explains why the error far from the 
plane is low even though the dust is only poorly measured. 
The fractional polarization outside P06 is typically only 1-2\%, 
where we use the degraded FDS dust map to trace intensity, and rises 
to $\sim$10\% in some regions of the plane. This is lower than the $\sim 4\%$ 
estimated in \citep{kogut/etal:2007,gold/etal:prep}. However, these maps 
are only estimated for $\nu \le 61$ GHz, and in regions of low 
dust the prior prefers zero polarization. The fractional
polarization estimate also assumes 
that the FDS map accurately traces the dust intensity in the \map\ frequency 
range. 
Inclusion of higher frequency data will allow us to learn more about
the polarized dust.

\section{Conclusion}
\label{sec:discuss}

We have used Bayesian parameter estimation to estimate low resolution
polarized CMB maps, marginalized over foreground contamination. 
These may then be used as inputs for a likelihood analysis. 
The emission model is parameterized accounting for physical 
understanding of the Galactic emission. 
The method has been tested on simulated maps, and found to produce unbiased estimates for the CMB power, 
quantified by the optical depth to reionization. 
With the five-year \map\ data we find a consistent result compared 
to template cleaning, with $\tau=0.090\pm0.019$ from this method and
$0.086\pm 0.016$ from the standard template-cleaning method. 
This method captures the increase 
in errors where foreground uncertainty
is larger, so depends less on a Galactic mask. 
Estimates of the polarized Galactic components indicate 
a synchrotron spectral index of order $\beta=-3.0$ in 
the Fan region in the Galactic anti-center, and the North Polar Spur area.

\acknowledgments
The \map\ mission is made possible by the support of the Science 
Mission Directorate Office at
NASA Headquarters.  This research was additionally 
supported by NASA grants NNG05GE76G,
NNX07AL75G S01, LTSA03-000-0090, ATPNNG04GK55G, and ADP03-0000-092.  
We thank the referee for helpful comments.
EK acknowledges support from an Alfred P. Sloan Research Fellowship.
This research has made 
use of NASA's Astrophysics Data System Bibliographic Services.  
We acknowledge use of the HEALPix, CAMB, and CMBFAST packages.


\begin{thebibliography}{51}
\expandafter\ifx\csname natexlab\endcsname\relax\def\natexlab#1{#1}\fi

\bibitem[{{Barnes} et~al.(2003){Barnes}, {Jefferys}, {Berger}, {Mueller},
  {Orr}, \& {Rodriguez}}]{barnes/etal:2003b}
{Barnes}, III, T.~G., {Jefferys}, W.~H., {Berger}, J.~O., {Mueller}, P.~J.,
  {Orr}, K., \& {Rodriguez}, R. 2003, \apj, 592, 539

\bibitem[{Beck(2001)}]{beck:2001}
Beck, R. 2001, Space Science Reviews, 99, 243, kluwer Academic Publishers

\bibitem[{{Bennett} et~al.(2003)}]{bennett/etal:2003}
{Bennett}, C.~L., et~al. 2003, \apj, 583, 1

\bibitem[{{Berdyugin} et~al.(2004){Berdyugin}, {Piirola}, \&
  {Teerikorpi}}]{berdyugin/piirola/teerikorpi:2004}
{Berdyugin}, A., {Piirola}, V., \& {Teerikorpi}, P. 2004, \aap, 424, 873

\bibitem[{{Berdyugin} et~al.(2001){Berdyugin}, {Teerikorpi}, {Haikala},
  {Hanski}, {Knude}, \& {Markkanen}}]{berdyugin/etal:2001}
{Berdyugin}, A., {Teerikorpi}, P., {Haikala}, L., {Hanski}, M., {Knude}, J., \&
  {Markkanen}, T. 2001, \aap, 372, 276

\bibitem[{{Crutcher} et~al.(2003){Crutcher}, {Heiles}, \&
  {Troland}}]{crutcher/heiles/troland:2003}
{Crutcher}, R., {Heiles}, C., \& {Troland}, T. 2003, in Lecture Notes in
  Physics, Berlin Springer Verlag, Vol. 614, Turbulence and Magnetic Fields in
  Astrophysics, ed. E.~{Falgarone} \& T.~{Passot},  155--181

\bibitem[{{Davis} \& {Greenstein}(1951)}]{davis/greenstein:1951}
{Davis}, L.~J. \& {Greenstein}, J.~L. 1951, \apj, 114, 206

\bibitem[{{Dobler} \& {Finkbeiner}(2007)}]{dobler/finkbeiner:2007}
{Dobler}, G. \& {Finkbeiner}, D.~P. 2007, ArXiv e-prints, 712

\bibitem[{{Draine}(2003)}]{draine:2003}
{Draine}, B.~T. 2003, \araa, 41, 241

\bibitem[{{Draine} \& {Fraisse}(2008)}]{draine/fraisse:prep}
{Draine}, B.~T. \& {Fraisse}, A.~A. 2008, ArXiv e-prints

\bibitem[{{Draine} \& {Lazarian}(1999)}]{draine/lazarian:1999}
{Draine}, B.~T. \& {Lazarian}, A. 1999, \apj, 512, 740

\bibitem[{{Draine} \& {Li}(2007)}]{draine/li:2007}
{Draine}, B.~T. \& {Li}, A. 2007, \apj, 657, 810

\bibitem[{Duncan et~al.(1995)Duncan, Haynes, Jones, \&
  Stewart}]{duncan/etal:1995}
Duncan, A.~R., Haynes, R.~F., Jones, K.~L., \& Stewart, R.~T. 1995, \mnras,
  277, 36

\bibitem[{{Dunkley} et~al.(2005){Dunkley}, {Bucher}, {Ferreira}, {Moodley}, \&
  {Skordis}}]{dunkley/etal:2005}
{Dunkley}, J., {Bucher}, M., {Ferreira}, P.~G., {Moodley}, K., \& {Skordis}, C.
  2005, \mnras, 356, 925

\bibitem[{{Dunkley} et~al.(2008)}]{dunkley/etal:prep}
{Dunkley}, J. et~al. 2008, \apjs

\bibitem[{{Eriksen} et~al.(2006)}]{eriksen/etal:2006}
{Eriksen}, H.~K., et~al. 2006, \apj, 641, 665

\bibitem[{{Eriksen} et~al.(2007)}]{eriksen/etal:2007}
---. 2007, \apj, 656, 641

\bibitem[{Finkbeiner et~al.(1999)Finkbeiner, Davis, \&
  Schlegel}]{finkbeiner/davis/schlegel:1999}
Finkbeiner, D.~P., Davis, M., \& Schlegel, D.~J. 1999, \apj, 524, 867

\bibitem[{{Gelfand} \& {Smith}(1990)}]{gelfand/smith:1990}
{Gelfand}, A.~E. \& {Smith}, A.~F.~M. 1990, Jour. Amer. Statist. Assn, 85, 398

\bibitem[{{Geweke} \& {Tanizaki}(2001)}]{geweke/tanizaki:2001}
{Geweke}, J. \& {Tanizaki}, H. 2001, Computational Statistics \& Data Analysis,
  37, 151

\bibitem[{{Gold} et~al.(2008)}]{gold/etal:prep}
{Gold}, B. et~al. 2008, \apjs

\bibitem[{Gorski et~al.(2005)Gorski, Hivon, Banday, Wandelt, Hansen, Reinecke,
  \& Bartlemann}]{gorski/etal:2004}
Gorski, K.~M., Hivon, E., Banday, A.~J., Wandelt, B.~D., Hansen, F.~K.,
  Reinecke, M., \& Bartlemann, M. 2005, \apj, 622, 759

\bibitem[{{Han}(2006)}]{han:2006b}
{Han}, J.-L. 2006, Chinese Journal of Astronony and Astrophysics, submitted
  (astro-ph/0603512)

\bibitem[{{Heiles}(2000)}]{heiles:2000}
{Heiles}, C. 2000, \aj, 119, 923

\bibitem[{{Hildebrand} \& {Dragovan}(1995)}]{hildebrand/dragovan:1995}
{Hildebrand}, R.~H. \& {Dragovan}, M. 1995, \apj, 450, 663

\bibitem[{{Hill} et~al.(2008)}]{hill/etal:prep}
{Hill}, R. et~al. 2008, \apjs

\bibitem[{{Hinshaw} et~al.(2007)}]{hinshaw/etal:2007}
{Hinshaw}, G., et~al. 2007, \apjs, 170, 288

\bibitem[{{Hinshaw} et~al.(2008)}]{hinshaw/etal:prep}
{Hinshaw}, G. et~al. 2008, \apjs

\bibitem[{{Jarosik} et~al.(2007)}]{jarosik/etal:2007}
{Jarosik}, N., et~al. 2007, \apjs, 170, 263


\bibitem[{{Jeffreys} (1961)}]{jeffreys:1961}
{Jeffreys}, H., 1961, {\it Theory of probability}, third edition, Oxford, 
Clarendon Press 

\bibitem[{{Kamionkowski} et~al.(1997){Kamionkowski}, {Kosowsky}, \&
  {Stebbins}}]{kamionkowski/kosowsky/stebbins:1997}
{Kamionkowski}, M., {Kosowsky}, A., \& {Stebbins}, A. 1997, \prd, 55, 7368

\bibitem[{{Knox} et~al.(2001){Knox}, {Christensen}, \&
  {Skordis}}]{knox/christensen/skordis:2001}
{Knox}, L., {Christensen}, N., \& {Skordis}, C. 2001, \apjl, 563, L95

\bibitem[{{Kogut} et~al.(2007)}]{kogut/etal:2007}
{Kogut}, A., et~al. 2007, \apj, 665, 355

\bibitem[{{Komatsu} et~al.(2008)}]{komatsu/etal:prep}
{Komatsu}, E. et~al. 2008, \apjs

\bibitem[{{Lawson} et~al.(1987){Lawson}, {Mayer}, {Osborne}, \&
  {Parkinson}}]{lawson/etal:1987}
{Lawson}, K.~D., {Mayer}, C.~J., {Osborne}, J.~L., \& {Parkinson}, M.~L. 1987,
  \mnras, 225, 307

\bibitem[{{Lewis} \& {Bridle}(2002)}]{lewis/bridle:2002}
{Lewis}, A. \& {Bridle}, S. 2002, \prd, 66, 103511

\bibitem[{{Miville-Deschenes} et~al.(2008){Miville-Deschenes}, {Ysard},
  {Lavabre}, {Ponthieu}, {Macias-Perez}, {Aumont}, \&
  {Bernard}}]{miville-deschenes/etal:prep}
{Miville-Deschenes}, M.~., {Ysard}, N., {Lavabre}, A., {Ponthieu}, N.,
  {Macias-Perez}, J.~F., {Aumont}, J., \& {Bernard}, J.~P. 2008, ArXiv
  e-prints, 802

\bibitem[{{Nolta} et~al.(2008)}]{nolta/etal:prep}
{Nolta}, M.~R. et~al. 2008, \apjs

\bibitem[{{Page} et~al.(2007)}]{page/etal:2007}
{Page}, L., et~al. 2007, \apjs, 170, 335

\bibitem[{{Reich} \& {Reich}(1988)}]{reich/reich:1988}
{Reich}, P. \& {Reich}, W. 1988, \aaps, 74, 7

\bibitem[{{Rybicki} \& {Lightman}(1979)}]{rybicki/lightman:1979}
{Rybicki}, G.~B. \& {Lightman}, A. 1979, {Radiative Processes in Astrophysics }
  (Wiley \& Sons: New York)

\bibitem[{{Seljak}(1997)}]{seljak:1997}
{Seljak}, U. 1997, \apj, 482, 6

\bibitem[{{Spergel} et~al.(2007)}]{spergel/etal:2007}
{Spergel}, D.~N., et~al. 2007, \apjs, 170, 377

\bibitem[{{Spitzer}(1998)}]{spitzer:1998}
{Spitzer}, L. 1998, {Physical Processes in the Interstellar Medium} (Physical
  Processes in the Interstellar Medium, by Lyman Spitzer, pp.~335.~ISBN
  0-471-29335-0.~Wiley-VCH , May 1998.)

\bibitem[{{Strong} et~al.(2007){Strong}, {Moskalenko}, \&
  {Ptuskin}}]{strong/moskalenko/ptuskin:2007}
{Strong}, A.~W., {Moskalenko}, I.~V., \& {Ptuskin}, V.~S. 2007, Annual Review
  of Nuclear and Particle Science, 57, 285

\bibitem[{{Strong} et~al.(2000){Strong}, {Moskalenko}, \&
  {Reimer}}]{strong/moskalenko/reimer:2000}
{Strong}, A.~W., {Moskalenko}, I.~V., \& {Reimer}, O. 2000, \apj, 537, 763

\bibitem[{{Uyaniker} et~al.(1999){Uyaniker}, {F{\"u}rst}, {Reich}, {Reich}, \&
  {Wielebinski}}]{uyaniker/etal:1999}
{Uyaniker}, B., {F{\"u}rst}, E., {Reich}, W., {Reich}, P., \& {Wielebinski}, R.
  1999, \aaps, 138, 31

\bibitem[{{Vall{\'e}e}(2005)}]{vallee:2005}
{Vall{\'e}e}, J.~P. 2005, \apj, 619, 297

\bibitem[{{Wandelt} et~al.(2004){Wandelt}, {Larson}, \&
  {Lakshminarayanan}}]{wandelt/larson/lakshminarayanan:2004}
{Wandelt}, B.~D., {Larson}, D.~L., \& {Lakshminarayanan}, A. 2004, \prd, 70,
  083511

\bibitem[{{Wolleben} et~al.(2006){Wolleben}, {Landecker}, {Reich}, \&
  {Wielebinski}}]{wolleben/etal:2006}
{Wolleben}, M., {Landecker}, T.~L., {Reich}, W., \& {Wielebinski}, R. 2006,
  \aap, 448, 411

\bibitem[{{Wright} et~al.(2008)}]{wright/etal:prep}
{Wright}, E.~L. et~al. 2008, \apjs

\bibitem[{{Zaldarriaga} \& {Seljak}(1997)}]{zaldarriaga/seljak:1997}
{Zaldarriaga}, M. \& {Seljak}, U. 1997, \prd, 55, 1830

\end{thebibliography}
\end{document}